\documentclass[reprint,twocolumn,groupedaddress,showpacs, longbibliography]{revtex4-1}

\usepackage{amsmath,amssymb,amsfonts}
\usepackage{bm, latexsym}
\usepackage[dvipdfmx]{graphicx}
\usepackage{bbm,times}
\usepackage[normalem]{ulem}
\usepackage[usenames,dvipsnames]{color}
\usepackage[pdftex]{hyperref}

\begin{document}

\title{Universal description of dissipative Tomonaga-Luttinger liquids with SU($N$) spin symmetry: \\
Exact spectrum and critical exponents}
\author{Kazuki Yamamoto}
\email{yamamoto.kazuki.72n@st.kyoto-u.ac.jp}
\author{Norio Kawakami}
\affiliation{Department of Physics, Kyoto University, Kyoto 606-8502, Japan}

\date{\today}

\begin{abstract}
Universal scaling relations for dissipative Tomonaga-Luttinger (TL) liquids with SU($N$) spin symmetry are obtained for both fermions and bosons, by using asymptotic Bethe-ansatz solutions and conformal field theory (CFT) in one-dimensional non-Hermitian quantum many-body systems with SU($N$) symmetry. We uncover that the spectrum of dissipative TL liquids with SU($N$) spin symmetry is described by the sum of one charge mode characterized by a complex generalization of $c=1$ U(1) Gaussian CFT, and $N-1$ spin modes characterized by level-$1$ SU($N$) Kac-Moody algebra with the conformal anomaly $c=N-1$, and thereby dissipation only affects the charge mode as a result of spin-charge separation in one-dimensional non-Hermitian quantum systems. The derivation is based on a complex generalization of Haldane's ideal-gas description, which is implemented by the SU($N$) Calogero-Sutherland model with inverse-square long-range interactions.
\end{abstract}

\maketitle

\section{Introduction}
Recent advances in the study of ultracold atoms have opened a new arena to investigate open quantum many-body systems, where a variety of unique phenomena that have no counterpart in isolated quantum systems occur \cite{Muller12, Daley14, Diehl16, Ashida20, Diehl08NP, Kraus08, Garcia09, Durr09, Yamamoto20, Yamamoto21}. Experimentally, high controllability of ultracold atoms has facilitated investigations of nonequilibrium quantum dynamics in open quantum systems, where strong coupling to the environment plays a vital role \cite{Syassen08, Yan13, Ott13, Rey14, Tomita17, Spon18, Tomita19, Gerbier20, Honda22}. In particular, non-Hermitian (NH) Hamiltonians have shed new light on investigations of dissipation-induced quantum many-body physics \cite{Nakagawa18, Louren18, Hanai20}; examples include loss-induced quantum phase transitions \cite{Yamamoto19, Hamazaki19, Hanai19}, measurement-induced entanglement dynamics \cite{Fisher18, Smith19, Nahum19, Fuji20, Goto20, Tang20, Buchhold21, Block21, Minato21}, NH quantum phases as a result of postselections of measurement outcomes by means of quantum-gas microscopy \cite{Ott16Rev, Ashida16, Ashida17, Nakagawa20, Yamamoto22}. Notably, one-dimensional (1D) NH quantum many-body systems show intriguing quantum critical phenomena induced by dissipation; e.g., correlation functions of dissipative Tomonaga-Luttinger (TL) liquids show unique critical behavior characterized by two TL parameters \cite{Ashida16, Yamamoto22}, and exceptional points, where the effective Hamiltonian cannot be diagonalized, cause anomalous singularities that accompany the divergence of correlation length \cite{Nakagawa21}. These studies have shown that dissipation drastically alters the universal properties of quantum many-body phenomena in isolated systems.

Another important aspect which has attracted broad interest in ultracold atoms is a multicomponent generalization of many-body phenomena, which show rich quantum phases as a result of strong correlation \cite{Wu03, Honerkamp04, Cherng07, Cazalilla09, Hermele09, Rey10, Cazalilla14, Capponi16, Yoshida21, Nakagawa22}. In particular, multicomponent fermions have been actively investigated in experiments over the last decade by controlling internal (nuclear) spin degrees of freedom, providing unique opportunities for quantum simulations of many-body systems. The milestone experiments have realized observations of antiferromagnetic correlations in SU($2$) Fermi-Hubbard models with degenerate ${}^{40}$K and ${}^6$Li \cite{Greif13, Hart15, Boll16, Parsons16, Cheuk16, Mazurenko17}, and exotic quantum phases stemming from SU($N$) ($N>2$) spin symmetry with ultracold alkaline-earth-like atoms such as ${}^{173}$Yb and ${}^{87}$Sr have recently been reported \cite{Taie12, Scazza14, Zhang14, Hofrichter16, Ozawa18, Tusi22, Taie22}. Multicomponent bosons are also successfully loaded into an optical lattice, where hyperfine states of bosonic atoms like ${}^{87}$Rb can be used, offering the possibility to explore novel highly entangled many-body states \cite{Yang20, Sun21}. Moreover, it is worth noting that physics with SU($N$) symmetry in ultracold atoms has recently extended its research area to nonequilibrium quantum systems as represented by SU($N$) Hubbard models with two-body loss \cite{Spon18, Yamamoto19, Nakagawa20, Yamamoto21, Nakagawa21, Rosso22, Rosso22arXiv}. Thus, it is natural to consider how the universal properties of many-body physics with SU($N$) symmetry are affected by dissipation. However, it is a highly nontrivial problem to identify the universality class of dissipative quantum many-body systems with internal degrees of freedom.

To study universal properties of strongly-correlated systems, 1D critical systems have been a subject of intense research in condensed matter physics, because they realize TL liquids, which present a general description of low-energy quantum many-body phenomena \cite{Haldane80, Haldane81C, Haldane81A, Haldane81}. Importantly, conformal invariance in 1D critical systems has brought about many valuable insights into many-body quantum systems \cite{BPZ84, BPZ84stat, Daniel84, Brezin88}, where TL liquids are characterized by U(1) Gaussian conformal field theory (CFT) with the central charge $c=1$. To access the universality class of TL liquids, Haldane has proposed that quantum models with inverse-square long-range interactions, which were initially introduced by Calogero and Sutherland \cite{Calogero69, Sutherland71a, Sutherland71b, Sutherland71c, Sutherland72}, give a unified understanding of 1D quantum critical phenomena \cite{Haldane88, Shastry88, Haldane91, Kawakami91, Kawakami94rev}. He has demonstrated that the model gives an ideal-gas description in one dimension, providing a variant notion of fractional statistics, dubbed fractional exclusion statistics \cite{Haldane91fra, Kawakami93}, which has an intimate relation to the fractional quantum Hall effect \cite{Laughlin83, Jain89}. For example, semion, a particle with statistical $1/2$ interactions, appears in the $S=1/2$ Heisenberg spin chain with inverse-square exchange \cite{Haldane88, Shastry88, Haldane91}. Studies with inverse-square interactions have been successfully generalized to multicomponent systems, and they have demonstrated the universal properties of TL liquids with SU($N$) spin symmetry \cite{Ha92, Kawakami92b, Kawakami92c, Kawakami93a, Kawakami93b, Bernard93}.

In this paper, we elucidate the universal properties of dissipative TL liquids with SU($N$) spin symmetry for both fermions and bosons in one dimension, based on a complex generalization of Haldane's ideal-gas description with the NH SU($N$) Calogero-Sutherland (CS) model. As a main result, we obtain critical exponents describing general NH quantum critical systems with SU($N$) symmetry. By analyzing the NH CS model, we obtain the asymptotic Bethe-ansatz (ABA) solutions and the finite-size scaling formula in CFT. We demonstrate that the spectrum of dissipative TL liquids with SU($N$) spin symmetry is described by one charge mode characterized by a complex generalization of $c=1$ U(1) Gaussian CFT, and $N-1$ spin modes characterized by level-$1$ SU($N$) Kac-Moody algebra with $c=N-1$. Since the spin modes are protected by SU($N$) symmetry, dissipation only affects the charge mode as a result of spin-charge separation in 1D NH quantum systems. Although our results are derived from specific integral models, universal scaling relations obtained can be applied to a wide range of NH quantum critical systems which are experimentally relevant, e.g., dissipative TL liquids with SU(2) spin symmetry describe the critical properties of the Fermi-Hubbard model with a complex-valued interaction.

The rest of this paper is organized as follows. First, in Sec.~\ref{sec_main_results}, we summarize the main results of the universal scaling relations in NH quantum critical systems for SU($N$) fermions and bosons, including SU(2) fermions as a special case. In the subsequent sections, we explain the detailed derivations of the universal properties of dissipative TL liquids with and without internal symmetry, based on a complex generalization of Haldane's ideal-gas description. In Sec.~\ref{sec_dissipativeTL}, we analyze the NH CS model based on the ABA solutions and the finite-size scaling analysis in CFT. We generalize the results to multicomponent systems by analyzing NH SU($N$) CS model in Sec.~\ref{sec_dissipativeSUNTL}. We finally conclude with a summary and outlook in Sec.~\ref{sec_discussion}.

\section{Main results: Universal scaling relations}
\label{sec_main_results}
In this section, we present a summary of the main results of this paper, that is, universal scaling relations for dissipative TL liquids with SU($N$) spin symmetry. The class of models that we want to consider throughout this paper is the critical system described by a NH Hamiltonian with SU($N$) symmetry. We note that SU($N$) symmetry is imposed not only on the original Hermitian Hamiltonian but also on dissipation, resulting in the whole SU($N$) symmetry of the NH Hamiltonian. Although the scaling relations are obtained by an ideal-gas approach based on specific integral models as detailed in the following sections \ref{sec_dissipativeTL} and \ref{sec_dissipativeSUNTL}, they are universal and applicable to generic 1D dissipative many-body systems.

Before going to the main results of scaling relations, we introduce important concepts of two-types of correlation functions that appear in NH systems. In NH systems that are described by the effective Hamiltonian $H_\mathrm{eff}$, a right eigenstate, which is defined by $H_\mathrm{eff}|\Psi^R\rangle=E|\Psi^R\rangle$, and a left eigenstate, which is defined by $H_\mathrm{eff}^\mathrm{\dag}|\Psi^L\rangle=E^*|\Psi^L\rangle$, are different from each other. Therefore, two types of correlation functions can emerge according to whether the right or left eigenstate is assigned to the bra vector in the expectation value. The first type is a mathematical generalization of correlation functions to NH systems defined by ${}_L\langle\cdots\rangle_R\equiv \langle\Psi_g^L|\cdots|\Psi_g^R\rangle/\langle\Psi_g^L|\Psi_g^R\rangle$, where $|\Psi_g^L\rangle$ and $|\Psi_g^R\rangle$ are the left and right ground states (in the sense of the real part of the energy) of $H_\mathrm{eff}$, respectively. This type of correlation functions is calculated through path integrals \cite{Yamamoto19, Yamamoto22}, and has been shown that it is directly related to a complex extension of CFT \cite{Yamamoto22}. The second type is defined by ${}_R\langle\cdots\rangle_R\equiv \langle\Psi_g^R|\cdots|\Psi_g^R\rangle/\langle\Psi_g^R|\Psi_g^R\rangle$, which is obtained in the postselected sector with no loss events as follows. First, the dynamics of the quantum state in such sectors is described by the Schr\"odinger equation $i\partial_t|\psi\rangle=H_{\mathrm {eff}}|\psi\rangle$. Then, with the use of the postselected ground state $|\psi\rangle=|\Psi_g^R\rangle$, we obtain a standard quantum-mechanical expectation value ${}_R\langle\cdots\rangle_R$, which corresponds to an experimentally measurable physical quantity \cite{Ashida16, Yamamoto22}. We call the correlation functions ${}_L\langle\cdots\rangle_R$ and ${}_R\langle\cdots\rangle_R$ the biorthogonal correlation function and the right-state correlation function, respectively. We note that the subscripts $L$ and $R$ for the brackets are not related to the left and right branches of TL liquids. In the following two subsections, we give universal scaling relations for right-state correlation functions, which are physical observables and relevant to cold-atom experiments.

\subsection{Critical exponents for fermions described by dissipative TL liquids with SU(2) spin symmetry}
We first focus on fermions described by dissipative TL liquids with SU(2) spin symmetry separately, since such SU(2) models have been by far the best studied until now both in solid state systems and cold atom systems. In particular, SU(2) fermions are relevant to Fermi-Hubbard models, which has been a subject of intense study in condensed matter physics.

Now, we give universal scaling relations for dissipative TL liquids with SU(2) spin symmetry. In NH fermionic quantum critical systems with SU(2) symmetry (with zero magnetic field), we obtain the long-distance behavior of the right-state charge-density correlator as
\begin{align}
&{}_R\langle\rho(x)\rho(0)\rangle_R\notag\\
&\simeq A_1 \cos(4k_Fx)x^{- \beta_1}+A_2 \cos(2k_Fx)x^{- \beta_2}+A_0\frac{1}{x^2},
\end{align}
where $\rho(x)=\sum_\sigma c_\sigma^\dag(x)c_\sigma(x)$, $A_j$ is a correlation amplitude, and the critical exponents $\beta_1$ and $\beta_2$ are given by
\begin{align}
\beta_1&=4K_\rho^\phi,\label{eq_beta1_SU2}\\
\beta_2&=1+K_\rho^\phi.
\label{eq_beta2_SU2}
\end{align}
As the ground state is given by the SU(2) singlet, we have defined the Fermi momentum $k_F=\pi n / 2$ with the density $n=M/L$, where $M$ is the total number of fermions and $L$ is the circumference of the system imposed by periodic boundary conditions. Importantly, the exponent $K_\rho^\phi$ is obtained by the complex-valued TL parameter $\tilde K_\rho$ as \cite{Yamamoto22}
\begin{align}
\frac{1}{K_\rho^\phi}=\mathrm{Re}\left[\frac{1}{\tilde K_\rho}\right],
\label{eq_Kphi}
\end{align}
where $\tilde K_\rho$ is defined through the $4k_F$ oscillating piece in the biorthogonal correlation function ${}_L\langle\rho(x)\rho(0)\rangle_R\simeq \cos(4k_Fx)x^{- \tilde \beta_1}$ as
\begin{align}
\tilde K_\rho=\frac{\tilde \beta_1}{4}.
\label{eq_complexTLparameterSU2}
\end{align}
We emphasize that the TL parameter is affected by dissipation, making it different from the one for the standard Hermitian TL liquid, and this fact leads to the emergence of two kinds of real TL exponents, which characterize experimentally measurable quantities as shown below. Here and henceforth, we use the symbol $\tilde A$ to emphasize that the quantity $\tilde A$ is complex-valued.

For the right-state fermion correlator, the long-distance behavior is written as
\begin{align}
{}_R\langle c_\sigma^\dag(x)c_\sigma(0)\rangle_R\simeq C_1 \cos(k_Fx)x^{-\eta_F},
\label{eq_correlator_fermion}
\end{align}
where $C_1$ is a correlation amplitude and the critical exponent $\eta_F$ is given by
\begin{align}
\eta_F=\frac{1}{2}+\frac{1}{4K_\rho^\theta}+\frac{K_\rho^\phi}{4}.
\label{eq_etaF_SU2}
\end{align}
The exponent $K_\rho^\theta$ is obtained from the complex-valued TL parameter $\tilde K_\rho$ as \cite{Yamamoto22}
\begin{align}
K_\rho^\theta = \mathrm{Re}[\tilde K_\rho].
\label{eq_Ktheta}
\end{align}
We note that Eq.~\eqref{eq_correlator_fermion} does not depend on spin indices for the SU(2) singlet ground state.

We emphasize that the scaling relations \eqref{eq_beta1_SU2}, \eqref{eq_beta2_SU2}, and \eqref{eq_etaF_SU2} are universal, and describe the general critical properties of fermions characterized by dissipative TL liquids with SU(2) spin symmetry. Eqs.~\eqref{eq_Kphi} and \eqref{eq_Ktheta} also highlight the universal properties of the U(1) charge part of dissipative TL liquids, which are characterized by the complex-valued TL parameter \cite{Yamamoto22}.

One of the prototypical dissipative models that have SU(2) symmetry is given by the NH Fermi-Hubbard model \cite{Nakagawa20, Nakagawa21, Yamamoto19, Yamamoto21, Rosso22arXiv}
\begin{equation}
H_\mathrm{eff}^\mathrm{Hubbard}=-t\sum_{j\sigma}(c_{j\sigma}^\dag c_{j+1\sigma}+\mathrm{H.c.})+\tilde U\sum_j n_{j\uparrow}n_{j\downarrow},
\label{eq_Hubbard}
\end{equation}
where $\tilde U$ is a complex-valued interaction as a result of dissipation, $t$ is a hopping parameter, $\sigma$ denotes up or down spin of fermions, $c_{j\sigma}$ is the annihilation operator of a spin-$\sigma$ fermion at site $j$, and $n_{j\sigma}\equiv c_{j\sigma}^\dag c_{j\sigma}$. Such model is realized by introducing two-body loss due to inelastic collisions between fermions as observed in cold-atom experiments \cite{Spon18, Honda22} e.g. with the use of photoassociation lasers. In the following, we consider dissipation that gives complex-valued interactions, and therefore other types of dissipation that cause e.g. asymmetric hopping of particles are not addressed in our study \cite{Fukui98}. We note that, in the previous studies that deal with the NH Fermi-Hubbard model, such as superfluid states with attractive interactions \cite{Yamamoto19, Yamamoto21} and Mott insulators with repulsive interactions \cite{Nakagawa20, Nakagawa21} have been considered, but the critical properties of metalic phases have not been investigated so far.

We demonstrate that critical properties of such NH SU(2) quantum models are described by dissipative TL liquids with SU(2) spin symmetry discussed in this subsection. In the NH Fermi-Hubbard model \eqref{eq_Hubbard}, critical properties of metalic phases are described by dissipative TL liquids with U(1)$\times$SU(2) symmetry, and the universal scaling relations are given by Eqs.~\eqref{eq_beta1_SU2}, \eqref{eq_beta2_SU2}, and \eqref{eq_etaF_SU2} by taking the continuum limit as $c_\sigma(x)=c_{j\sigma}/\sqrt{a}$ with the lattice constant $a$.

\subsection{Critical exponents for fermions and bosons described by dissipative TL liquids with SU($N$) spin symmetry}
In this subsection, motivated by the recent progress in SU($N$) quantum phenomena in ultracold atoms, we summarize a generalization of universal scaling relations in NH quantum critical systems to SU($N$) symmetric cases. We give universal scaling relations for both fermions and bosons described by dissipative TL liquids with SU($N$) spin symmetry. The right-state charge-density correlator is the same for both fermions and bosons, and the long-distance behavior reads
\begin{align}
{}_R\langle\rho(x)\rho(0)\rangle_R
\simeq\sum_{j=1}^N A_j \cos[2(N-j+1)k_Fx]x^{- \beta_j}+\frac{A_0}{x^2},
\end{align}
where $A_j$ is a correlation amplitude, and the critical exponent $\beta_j$ corresponding to the $2(N-j+1)k_F$ oscillation is given by
\begin{align}
\beta_j=\frac{2(N-j+1)(j-1)}{N}+\frac{2(N-j+1)^2}{N}K_\rho^\phi.
\label{eq_beta_SUN}
\end{align}
Here, we have defined the Fermi momentum $k_F=\pi n / N$, which reflects the SU($N$) singlet ground state. As in the SU(2) case, the critical exponent $K_\rho^\phi$ is obtained by the complex-valued TL parameter $\tilde K_\rho$ through Eq.~\eqref{eq_Kphi}. In this case, $\tilde K_\rho$ is generalized to the system with SU($N$) symmetry, resulting in \begin{align}
\tilde K_\rho=\frac{\tilde \beta_1}{2N},
\label{eq_complexTLparameterSUN}
\end{align}
which corresponds to the $2Nk_F$ oscillating piece of the biorthogonal correlation function ${}_L\langle\rho(x)\rho(0)\rangle_R\simeq \cos(2Nk_Fx)x^{- \tilde \beta_1}$.

On the other hand, the single-particle correlator is different between fermions and bosons. For fermions described by dissipative TL liquids with SU($N$) spin symmetry, we obtain the same correlator as Eq.~\eqref{eq_correlator_fermion}, where the critical exponent $\eta_F$ is generalized to SU($N$) symmetric cases;
\begin{align}
\eta_F=\frac{N-1}{N}+\frac{1}{2NK_\rho^\theta}+\frac{K_\rho^\phi}{2N}.
\label{eq_etaFSUN}
\end{align}
We note that the exponent $K_\rho^\theta$ is obtained from Eq.~\eqref{eq_Ktheta} with Eq.~\eqref{eq_complexTLparameterSUN}, and the fermion correlator \eqref{eq_correlator_fermion} is independent of spin indices for the SU($N$) singlet ground state as in the SU(2) symmetric cases. For the right-state boson correlator, we obtain the long-distance behavior as
\begin{align}
{}_R\langle b_\sigma^\dag(x)b_\sigma(0)\rangle_R\simeq B_1x^{- \eta_B},
\label{eq_correlator_boson}
\end{align}
where $b_\sigma (x)$ is the annihilation operator of bosons, $B_1$ is a correlation amplitude, and the critical exponent $\eta_B$ is given by
\begin{align}
\eta_B=\frac{N-1}{2N}+\frac{1}{2NK_\rho^\theta}.
\label{eq_etaBSUN}
\end{align}
We note that the bosonic correlator \eqref{eq_correlator_boson} does not depend on the spin $\sigma$ as in the case of fermions. It is worth noting that the bosonic correlator \eqref{eq_correlator_boson} is not accompanied by the $k_F$ oscillation that appears in the fermion case because of Bose statistics (see Sec.~\ref{sec_dissipativeTL} for details).

We emphasize that the scaling relations \eqref{eq_beta_SUN}, \eqref{eq_etaFSUN}, and \eqref{eq_etaBSUN} are universal, and characterize the general critical properties of fermions and bosons described by dissipative TL liquids with SU($N$) spin symmetry, where the critical theory consists of one charge mode characterized by a complex generalization of $c=1$ U(1) Gaussian CFT \cite{Yamamoto22}, and $N-1$ spin modes characterized by $c=N-1$ level-1 SU($N$) Kac-Moody algebra (see Sec.~\ref{sec_dissipativeSUNTL} for details). Importantly, the algebraic structure of the spin sectors is fixed to the real one as a result of symmetry enhancement. In Sec.~\ref{sec_dissipativeSUNTL}, we give the detailed derivation which includes the ABA solutions and the finite-size scaling analysis in CFT based on a complex generalization of Haldane's ideal-gas description.

\section{Dissipative Tomonaga-Luttinger liquids without internal symmetry}
\label{sec_dissipativeTL}
In this section, we analyze the NH CS model without internal symmetry, based on a complex generalization of Haldane's ideal-gas description, which is embodied by the ABA solution. We will see that the ideal-gas description captures universal properties of single-component dissipative TL liquids.

\subsection{Model}
We consider a 1D long-range interacting NH quantum system with the circumference $L$ described by
\begin{gather}
H_\mathrm {eff} = - \sum_{j=1}^M\frac{\partial^2}{\partial x_j^2} + \sum_{j>l}V(x_j - x_l),\label{eq_Heff}\\
V(x)=\tilde g \sum_{n=-\infty}^\infty(x+nL)^{-2}=\frac{\tilde g \pi^2}{L^2}\left[\sin\left(\frac{\pi x}{L}\right)\right]^{-2},\label{eq_interaction}
\end{gather}
where $M$ is the number of particles, and $\tilde g$ is the dimensionless complex parameter of the inverse-square long-range interaction. This model is a generalization of the CS model \cite{Calogero69, Sutherland71a, Sutherland71b, Sutherland71c, Sutherland72, Kawakami91} to NH quantum many-body systems \cite{Footnote}.
Models with inverse-square interactions have been investigated intensively to date, for example, in a quantum spin chain (Haldane-Shastry model \cite{Haldane88, Shastry88, Haldane91}), in a supersymmetric $t-J$ model  \cite{Kuramoto91, Kawakami92a, Kawakami92b}, and in a continuum model interacting by the inverse-square potential (CS model). 
These models are known to be integrable \cite{Hikami93a, Hikami93b, Hikami93c}, for which the ground-state wave function is given by a Jastrow form, i.e., products of two-body wave functions \cite{Calogero69, Sutherland71a, Sutherland71b, Sutherland71c, Sutherland72}. 

An important aspect of this class of integrable models is that they can describe the essential properties of 1D quantum critical systems as an ideal gas, as demonstrated by Haldane \cite{Haldane88, Haldane91, Haldane91fra}. To explain the essence of the idea, let us start by analyzing the ground state of the NH CS model \eqref{eq_Heff}. A little algebra leads to a complex generalization of the ground-state wave function of the Jastrow form (see Appendix~\ref{sec_Jastrow} for detailed calculations)
\begin{align}
\Psi_g=\prod_{j>l}\left|\sin\frac{\pi(x_j-x_l)}{L}\right|^{\tilde \lambda-s}&\left(\sin\frac{\pi(x_j-x_l)}{L}\right)^{s},\label{eq_Jastrow}
\end{align}
where $x_j>x_l$, $\tilde \lambda =[(1+2\tilde g)^{\frac{1}{2}}+1]/2$, $\mathrm{Re}[\tilde \lambda]\ge1/2$ is assumed, and we have introduced $s=1$ for fermions and $s=0$ for bosons, respectively. The wave function \eqref{eq_Jastrow} reduces to the one for free fermions or hard-core bosons in the limit $\tilde \lambda\to 1$. The Jastrow-type wave function indicates that the two-body scattering is essential to the system. This observation motivated Sutherland to propose the ABA method \cite{Sutherland71a, Sutherland71b, Sutherland71c, Sutherland72}, which states that the many-body scattering is decomposed into the product of two-body scatterings as a result of the integrability in spite of long-range nature of interactions \cite{Sutherland04}. The ABA solution provides a concrete way to implement the ideal-gas description, as described below.

\subsection{Asymptotic Bethe ansatz: Ideal-gas description}
\begin{figure}[t]
\centering
\includegraphics[width=8cm]{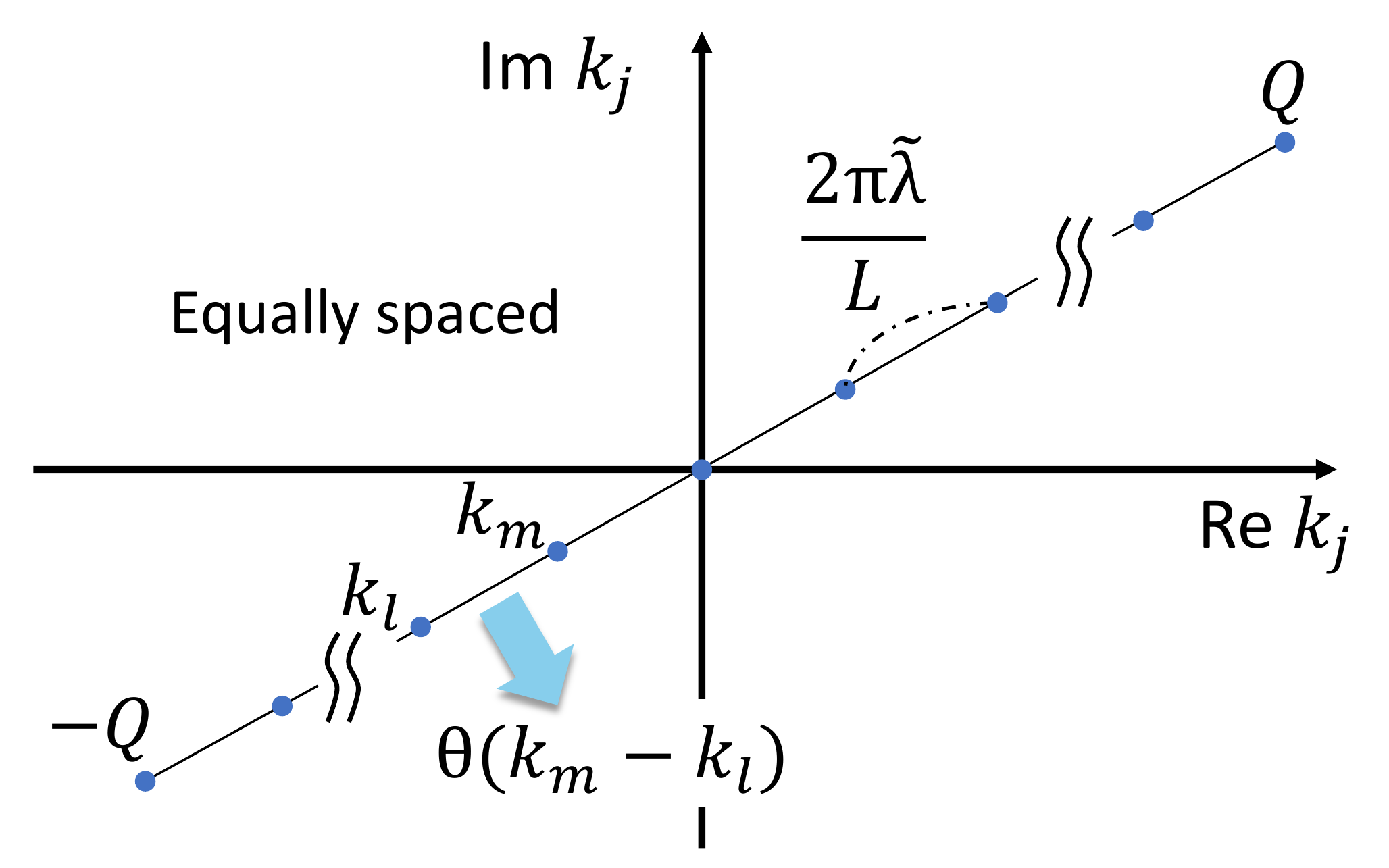}
\caption{Distribution of the quasimomentum $k_j$ in the ground state of long-range interacting systems \eqref{eq_Heff}. The quasimomentum $k_j$ is equally spaced along the straight line with the interval $2\pi\tilde\lambda/L$, and $\pm Q$, which are called the quasi Fermi points, denote the edges of the quasimomentum distribution. The generalized signature between $k_m$ and $k_l$ gives the phase shift $\theta(k_m-k_l)$ (see text).}
\label{fig_phaseshift}
\end{figure}

Now, we derive the ABA solution of the NH CS model. The phase shift as a result of the two-body scattering is read off from the Jastrow wave function \eqref{eq_Jastrow} as \cite{Kawakami91}
\begin{align}
\theta(k)=\pi (\tilde \lambda -1) \mathrm{sgn}^*(k),
\label{eq_phaseshift}
\end{align}
where we have defined the function $\mathrm{sgn}^*(k)$ for the complex quasimomentum $k=\alpha e^{i\xi}~(\alpha\in\mathbb{R}, -\pi/2<\xi<\pi/2)$ as $\mathrm{sgn}^*(k)=\mathrm{sgn}(\alpha)$ (see Fig.~\ref{fig_phaseshift}). 
There is an important feature in Eq.~\eqref{eq_phaseshift}: the phase shift is characterized by a single $k$-independent parameter $\tilde \lambda$ and a step-function in $k$ space, which is contrasted to ordinary integrable models for which the corresponding phase shift is $k$-dependent. We therefore see that the above two-body phase shift incorporates the interaction effect via a level repulsion parameter $\tilde \lambda$ in a way similar to ideal gases such as free fermions or hard-core bosons \cite{Haldane88, Haldane91, Haldane91fra}. 

The philosophy of the ABA (ideal-gas description) lies in the fact that the two-body phase shift \eqref{eq_phaseshift} exactly describes the whole spectrum of the long-range interacting system \eqref{eq_Heff} for arbitrary densities due to the integrability \cite{Sutherland04}. Based on this idea, we obtain the ABA equation as
\begin{gather}
k_jL=2\pi I_j +\sum_{i=1}^M\theta(k_j-k_i),
\label{eq_bethe}
\end{gather}
where 
\begin{align}
I_j:~\mathrm{integer},
\end{align}
for fermions, and 
\begin{align}
I_j=\frac{M+1}{2}~(\mathrm{mod}~1),
\end{align}
for bosons, respectively. It is worth noting that both fermions and bosons are described by the transcendental equation \eqref{eq_bethe}, and only the selection rule of the quantum number $I_j$ distinguishes their statistics. By using the quasimomentum $k_j$, we obtain the energy as $E=\sum k_j^2$, and the momentum as $P=\sum k_j$. We note that, in NH systems, integrability has been shown to hold by explicitly constructing the Bethe equation for models having e.g. asymmetric hoppings and complex-valued interactions \cite{Fukui98, Nakagawa18, Shibata19, Nakagawa21, Buca20, Yamamoto22}.

We consider the ground state and obtain the quasimomentum distribution. The quantum numbers for the ground state $\{I_j\}$ are successive integers (half-odd integers), which are distributed symmetrically around the origin. Then, we obtain the quasimomentum as
\begin{gather}
k_jL=\pi\tilde \lambda (2j-M-1)
\label{eq_kj_boson}
\end{gather}
for odd-$M$ fermions and bosons by using $I_j=(2j-M-1)/2$, and 
\begin{gather}
k_jL=\pi\tilde \lambda (2j-M-1)+\pi
\label{eq_kj_fermioneven}
\end{gather}
for even-$M$ fermions with the use of $I_j=(2j-M-1)/2+1/2$. We see from Eqs.~\eqref{eq_kj_boson} and \eqref{eq_kj_fermioneven} that the quasimomentum is equally spaced along the straight line in the complex plane as depicted in Fig.~\ref{fig_phaseshift}. This is nothing but the fact that each particle feels the complex-valued repulsion due to a step function in the complex phase shift given in Eq.~\eqref{eq_phaseshift}. As shown in the following, the constant shift $1/2$ in the even-$M$ fermion's quantum number $I_j$ leads to the characteristic selection rule for fermions, which convey the current accompanied by the change of the particle number. It is worth noting that bosons do not know the Fermi point as a result of Bose statistics, and this fact causes the difference between the single-particle correlators of fermions and bosons.

\subsection{Velocity of excitations}
Next, we calculate the velocity of excitations that controls the bulk quantities, for which we first obtain the dressed energy and the distribution function in the thermodynamic limit. We note that the velocity of excitations becomes complex reflecting the effect of dissipation in open quantum systems. Here, we consider an elementary excitation from the ground state. By inserting a hole into the ground-state distribution of the quasimomentum, the dressed energy $\epsilon(k)$ is calculated by using the ABA equation \eqref{eq_bethe} as
\begin{align}
\epsilon(k)=k^2-\tilde \mu-(\tilde \lambda-1)\int_{-Q}^Qdk^\prime\delta(k-k^\prime)\epsilon(k^\prime),
\end{align}
which leads to
\begin{align}
\epsilon(k)=
\begin{cases}
k^2-\tilde \mu&(|k|>|Q|),\\
\frac{1}{\tilde \lambda}(k^2-\tilde \mu)&(|k|<|Q|),
\end{cases}
\label{eq_dressedenergy}
\end{align}
where $Q=\pi n \tilde \lambda$ with the density $n=M/L$. Here, we have introduced the delta function $\delta(k)$ that satisfies $\int_C dk \delta(k-k_0)f(k)=f(k_0)$ for an arbitrary function $f(k)$ on the contour $C$, which is the straight line shown in Fig.~\ref{fig_phaseshift}. As the dressed energy satisfies $\epsilon(\pm Q)=0$ at the quasi Fermi points $\pm Q$, the complex-valued chemical potential is given by $\tilde \mu=\pi^2n^2\tilde \lambda^2$. Similarly, the distribution function in the thermodynamic limit $\sigma(k|\pm Q)$ is calculated as
\begin{align}
\sigma(k|\pm Q)=\frac{1}{2\pi}-(\tilde \lambda-1)\int_{-Q}^Qdk^\prime \delta(k-k^\prime)\sigma(k^\prime|\pm Q),
\end{align}
which leads to
\begin{align}
\sigma(k|\pm Q)=
\begin{cases}
\frac{1}{2\pi}&(|k|>|Q|),\\
\frac{1}{2\pi\tilde \lambda}&(|k|<|Q|).
\end{cases}
\label{eq_distributionfunction}
\end{align}
Finally, by using Eqs.~\eqref{eq_dressedenergy} and \eqref{eq_distributionfunction}, we obtain the velocity of excitations as
\begin{align}
\tilde v =\frac{\epsilon^\prime(Q)}{2\pi\sigma(Q|\pm Q)}=2\pi\tilde \lambda n,
\label{eq_velocity}
\end{align}
where $\epsilon^\prime(k)$ denotes the derivative of $\epsilon(k)$ with respect to $k$. We emphasize that the velocity of excitations \eqref{eq_velocity} becomes complex as a result of dissipation.

\subsection{Finite-size scaling analysis in CFT}
In this subsection, we perform the finite-size scaling analysis in CFT \cite{Cardy84a, Cardy84b, Cardy86, Cardy86log, Blote86, Affleck86} by explicitly calculating the ground-state energy, the excitation energy, and the momentum transfer for both fermions and bosons \cite{Kawakami91}. First, we calculate the ground-state energy. By substituting Eqs.~\eqref{eq_kj_boson} and \eqref{eq_kj_fermioneven} into $E=\sum k_j^2$, we arrive at the ground state energy for fermions and bosons as
\begin{align}
E_0=\frac{\pi^2{\tilde \lambda}^2}{L^2}\sum_{j=1}^M(2j-M-1)^2=L\epsilon_0-\frac{\pi\tilde v \tilde \lambda}{6L},
\label{eq_groundenergy}
\end{align}
where we have introduced $\epsilon_0=\pi^2\tilde\lambda^2n^3/3$, and the energy becomes complex as the system decays due to dissipation. We note that the calculation in Eq.~\eqref{eq_groundenergy} is exact, and we have shifted the ground-state energy for even-$M$ fermions by $\pi^2n/L$ because this term does not affect the universality. One notices that Eq.~\eqref{eq_groundenergy} is peculiar since the central charge could be given by $c=\tilde \lambda$, hence changes continuously in the complex plane. However, this artifact solely comes from the long-range nature of the interaction in a finite-size system, where we have applied the cylindrical geometry to the long-range interacting system \cite{Kawakami91}, and the universality class is described by the central charge $c=1$ as we see from the conformal dimensions below.

Next, we calculate the excitation energy from the ground state $\Delta E$ and the associated momentum transfer $P$. There are two types of excitations characterized by the quantum numbers $\Delta M$ and $\Delta D$ other than the particle-hole excitations. $\Delta M$ denotes the change of the particle number and $\Delta D$ describes the number of particles which move from the left Fermi point to the right one accompanied by the large momentum transfer $P$. Then, the excitation energy $\Delta E$ is calculated as
\begin{align}
\Delta E&=\frac{\pi^2{\tilde \lambda}^2}{L^2}\Bigg[\sum_{j=1}^{M+\Delta M}\left(2j-M-\Delta M-1+\frac{2\Delta D}{\tilde \lambda}\right)^2\notag\\
&\hspace{1.8cm}-\sum_{j=1}^M(2j-M-1)^2\Bigg]\notag\\
&=\tilde \mu \Delta M + \frac{2\pi\tilde v}{L}\left[\frac{\tilde \lambda}{4}(\Delta M)^2+\frac{1}{\tilde \lambda}(\Delta D)^2\right]+O\left(\frac{1}{L^2}\right).
\end{align}
The associated momentum transfer $P$ is obtained as
\begin{align}
P&=\frac{\pi \tilde \lambda}{L}\sum_{j=1}^{M+\Delta M}\left(2j-M-\Delta M -1 +\frac{2\Delta D}{\tilde \lambda}\right)\notag\\
&=2 k_F\Delta D+\frac{2\pi}{L}\Delta M \Delta D,
\end{align}
where $k_F=\pi n$ is the Fermi momentum. Importantly, the difference between fermions and bosons only appears in the selection rule between $\Delta M$ and $\Delta D$;
\begin{gather}
\Delta M:~\mathrm{integer},\label{eq_deltaM}\\
\Delta D=\frac{\Delta M}{2}~(\mathrm{mod}~1),
\end{gather}
for the Fermi case, and
\begin{gather}
\Delta M:\mathrm{integer},\\
\Delta D:\mathrm{integer},\label{eq_deltaD}
\end{gather}
for the Bose case, respectively. In the Fermi case, we cannot take the current $\Delta D$ to be independent of the change of the particle number $\Delta M$. Finally, by adding particle-hole excitation terms characterized by the quantum numbers $n^\pm$, which are nonnegative integers, we arrive at the finite-size scaling formula for the excitation energy and the associated momentum transfer as
\begin{gather}
\Delta E=\tilde \mu \Delta M + \frac{2\pi\tilde v}{L}\left[\frac{\tilde \lambda}{4}(\Delta M)^2+\frac{1}{\tilde \lambda}(\Delta D)^2+n^++n^-\right],\label{eq_excitationenergy}\\
P=2 k_F\Delta D+\frac{2\pi}{L}(\Delta M \Delta D+n^+-n^-).\label{eq_momentum}
\end{gather}
It is worth noting that the excitation energy becomes complex due to dissipation, but the momentum transfer remains real. From Eqs.~\eqref{eq_excitationenergy} and \eqref{eq_momentum}, we can read off the conformal weights characterizing the holomorphic and antiholomorphic parts of the underlying Virasoro algebra as
\begin{align}
\Delta^\pm (\Delta M;\Delta D;n^\pm)=\frac{1}{2}\left(\frac{\Delta M\sqrt{\tilde \lambda}}{2}\pm\frac{\Delta D}{\sqrt{\tilde \lambda}}\right)^2+n^\pm.\label{eq_conformalweight}
\end{align}
The conformal dimensions \eqref{eq_conformalweight} are typical for a complex generalization of $c=1$ U(1) Gaussian CFT \cite{Yamamoto22}, and the universal behavior of the system is characterized by the complex-valued TL parameter $\tilde K=1/\tilde \lambda$. The critical exponents of dissipative TL liquids have already been obtained in Ref.~\cite{Yamamoto22}, where we have calculated both biorthogonal correlation functions and right-state correlation functions, by using path integrals and ground-state wave function approach, respectively. Summarizing this section, we have confirmed that an ideal-gas description based on the ABA solution of the NH CS model captures the essential properties of dissipative TL liquids in NH quantum critical systems.

\section{Dissipative Tomonaga-Luttinger liquids with SU($N$) spin symmetry}
\label{sec_dissipativeSUNTL}
In this section, we generalize the NH CS model to multicomponent systems with SU($N$) symmetry, and analyze the NH SU($N$) CS model based on a complex generalization of Haldane's ideal-gas description.

\subsection{Model}
We consider a complex generalization of the SU($N$) CS model to 1D NH quantum systems. The NH SU($N$) CS model with the circumference $L$ is given by
\begin{align}
H_\mathrm{eff}=-\frac{1}{2}\sum_{i=1}^M\frac{\partial^2}{\partial x_i^2}+\sum_{i<j}D(x_i-x_j)^{-2}\tilde \lambda^\prime(\tilde \lambda^\prime +P_{ij}^\sigma),
\label{eq_HeffSUN}
\end{align}
where $D(x)=(L/\pi)\left|\sin\left({\pi x}/{L}\right)\right|$ is the chord distance, $P_{ij}^\sigma~(\sigma=1,2,\cdots,N)$ is an operator that exchanges particle spins between the sites $i$ and $j$, $\tilde \lambda^\prime$ is the dimensionless complex parameter of the long-range interaction, and we have assumed $\mathrm{Re}[\tilde \lambda^\prime]>0$. We note that the SU($N$) CS model was originally solved by Ha and Haldane \cite{Ha92}, and we use similar notations which are used in Ref.~\cite{Ha92}; $\tilde \lambda^\prime$ in Eq.~\eqref{eq_HeffSUN} and $\tilde \lambda$ in Eq.~\eqref{eq_Heff} are related by $\tilde \lambda^\prime=\tilde \lambda -1$, and Eqs.~\eqref{eq_HeffSUN} and \eqref{eq_Heff} are different from each other by a factor 2 in the single-component limit. Such difference in notations does not affect the universality.

The ground-state wave function of the SU($N$) CS model is a Jastrow type, and it has been demonstrated that the Jastrow factor does not contain the internal spin degrees of freedom \cite{Ha92, Kawakami92b, Kawakami92c, Kawakami93a}. Then, we obtain the ground-state wave function by generalizing the results obtained in the Hermitian case \cite{Ha92} to the NH case as
\begin{align}
\Psi_g=\prod_{l>m}|z_l-z_m|^{\tilde \lambda^\prime-s^\prime}\Psi_0,
\label{eq_JastrowSUN}
\end{align}
where
\begin{align}
\Psi_0=\prod_jz_j^{J_{\sigma_j}}\prod_{l>m}\Bigg\{&(z_l-z_m)^{s^\prime+\delta_{\sigma_l,\sigma_m}}\notag\\
&\times\exp\left[\frac{i}{2}\pi\mathrm{sgn}(\sigma_l-\sigma_m)\right]\Bigg\}.
\label{eq_psiSUN}
\end{align}
Here, $z_m=\exp(2\pi i x_m/L)$, $\sigma_l$ is the ordered spin index, $J_\sigma$ is the global current of the spin-$\sigma$ particle, $\delta$ is the Kronecker's delta, and we have introduced $s^\prime=0$ for fermions and $s^\prime=1$ for bosons, respectively. Thus, we see from Eq.~\eqref{eq_JastrowSUN} that the complex interaction parameter $\tilde \lambda^\prime$ in the Jastrow factor only affects the charge degrees of freedom even in the NH case.

\subsection{Nested asymptotic Bethe ansatz}
\begin{figure}[b]
\centering
\includegraphics[width=8.5cm]{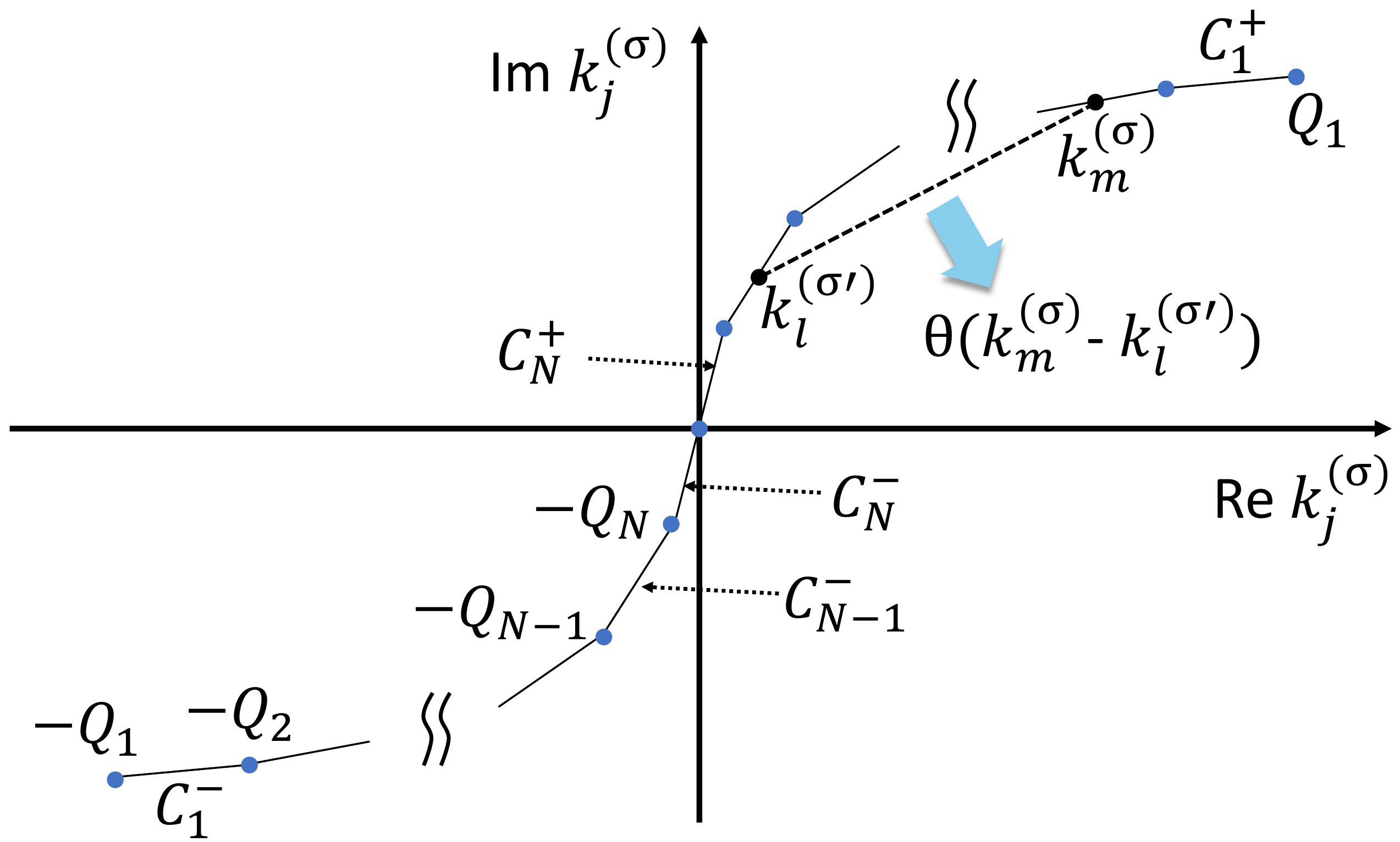}
\caption{Distribution of the quasimomentum $k_j^{(\sigma)}$ in the ground state of the NH SU($N$) CS model \eqref{eq_HeffSUN} (Chemical potential imbalance between different spin-$\sigma$ particles is included). Each region $C_\tau$ is constructed from two straight lines which are symmetric with respect to the origin; $C_\tau^+$ (the region between $Q_\tau$ and $Q_{\tau+1}$) and $C_\tau^-$ (the region between $-Q_\tau$ and $-Q_{\tau+1}$). Quasimomentum is equally spaced in each region $C_\tau$, and $k_j^{(\sigma)}$ is distributed in the region $\sum_{\tau=\sigma}^N C_\tau$, where $\pm Q_\sigma$ denote the endpoints of the quasimomentum distribution of $k_j^{(\sigma)}$. The generalized signature between $k_m^{(\sigma)}$ and $k_l^{(\sigma^\prime)}$ gives the phase shift $\theta(k_m^{(\sigma)}-k_l^{(\sigma^\prime)})$ (see text).}\label{fig_phaseshiftSUN}
\end{figure}

Next, to exploit an ideal-gas description, we derive the ABA solution of the NH SU($N$) CS model. By generalizing the $S$-matrices in the Hermitian multicomponent systems \cite{Yang67, Sutherland68, Kawakami92b, Sutherland75, Kawakami93a}, we obtain the $S$-matrices of the NH SU($N$) CS model as
\begin{gather}
S_{ij}=\exp(-i\tilde \lambda^\prime\theta(k_i-k_j))\lim_{\eta\to 0}\frac{k_i-k_j+i\eta P_{ij}^\sigma}{k_i-k_j+i\eta},\label{eq_SijSUNfermion}
\end{gather}
for fermions, and 
\begin{align}
S_{ij}=\exp(-i(\tilde \lambda^\prime-1)\theta(k_i-k_j))\lim_{\eta\to 0}\frac{k_i-k_j+i\eta P_{ij}^\sigma}{k_i-k_j-i\eta},\label{eq_SijSUNboson}
\end{align}
for bosons, respectively. Here, we have introduced $\theta(k)=\pi\mathrm{sgn}^*(k)$, where the function $\mathrm{sgn}^*(k)$ for the complex quasimomentum $k=\alpha e^{i\xi}~(\alpha\in\mathbb{R},-\pi/2<\xi<\pi/2)$ is defined by $\mathrm{sgn}^*(k)=\mathrm{sgn}(\alpha)$ (see Fig.~\ref{fig_phaseshiftSUN}) in the same way as in the case without internal symmetry. It is noted that, for $k_i-k_j=\alpha e^{i\xi}$ in the $S$-matrices, we take the limit $\eta\to0$ as $\eta^\prime\to+0$ with $\eta=\eta^\prime e^{i\xi}~(\eta^\prime\in\mathbb{R})$. The additional phase shift in the exponential in the $S$-matrices $\pi(\tilde \lambda^\prime - s^\prime)\mathrm{sgn}^*(k_i-k_j)$ comes from the Jastrow factor in Eq.~\eqref{eq_JastrowSUN}. The $S$-matrices in Eqs.~\eqref{eq_SijSUNfermion} and \eqref{eq_SijSUNboson} satisfy the Yang-Baxter equation $S_{jk}S_{ik}S_{ij}=S_{ij}S_{ik}S_{jk}$, hence we can construct the nested ABA \cite{Yang67, Sutherland68, Kawakami92b, Sutherland75, Kawakami93a} in NH multicomponent systems with SU($N$) symmetry. We note that, in the limit $\tilde\lambda^\prime\to0$, the $S$-matrix for fermions \eqref{eq_SijSUNfermion} reduces to the one for free fermions with SU($N$) internal spin symmetry \cite{Yang67, Sutherland68}. As for the limit $\tilde\lambda^\prime-1\to0$, the $S$-matrix for bosons \eqref{eq_SijSUNboson} corresponds to the one for the SU($N$) Haldane-Shastry model \cite{Kawakami92b, Sutherland75}.

By using the $S$-matrices~\eqref{eq_SijSUNfermion} and \eqref{eq_SijSUNboson}, the nested ABA equations for fermions ($s^\prime=0$) and bosons ($s^\prime=1$) are calculated as \cite{Yang67, Sutherland68, Kawakami92b, Sutherland75, Kawakami93a}
\begin{widetext}
\begin{gather}
\exp(ik_j^{(1)}L)=(-1)^{(M_1-1)s^\prime}\prod_m^{M_2}f_1(k_m^{(2)}-k_j^{(1)})\prod_{l(\neq j)}^{M_1}\exp\left[-i\tilde \lambda^\prime\theta(k_l^{(1)}-k_j^{(1)})\right],\quad(1\le j\le M_1),\label{eq_SUNCS1}\\
\prod_{l(\neq m)}^{M_\sigma}f_2(k_m^{(\sigma)}-k_l^{(\sigma)})=\prod_j^{M_{\sigma-1}}f_1(k_m^{(\sigma)}-k_j^{(\sigma-1)})\prod_{q}^{M_{\sigma+1}}f_1(k_m^{(\sigma)}-k_q^{(\sigma+1)}),\quad(1\le m\le M_\sigma,~2\le\sigma\le N-1),\label{eq_SUNCS2}\\
\prod_{l(\neq q)}^{M_{N}}f_2(k_q^{(N)}-k_l^{(N)})=\prod_j^{M_{N-1}}f_1(k_q^{(N)}-k_j^{(N-1)}),\quad(1\le q\le M_{N}),\label{eq_SUNCS3}
\end{gather}
respectively. Here, we have introduced $f_m(x)=\lim_{\eta\to0}(x-im\eta/2)/(x+im\eta/2)$, and the number of the quasimomentum $k_j^{(\sigma)}$ as $M_\sigma=\sum_{\tau=\sigma}^N N_\tau$, where $N_\sigma$ is the number of particles with spin $\sigma$. The nested ABA equations \eqref{eq_SUNCS1}--\eqref{eq_SUNCS3} describe a charge excitation ($\sigma=1$) characterized by the complex-valued interaction parameter $\tilde \lambda^\prime$, and $N-1$ kinds of spin excitations ($2\le\sigma \le N$), where $\tilde \lambda^\prime$ does not appear. The particles that appear in the spin sectors are called spinons, which are frequently used to describe spin excitations in the Haldane-Shastry model \cite{Haldane91}. We see that the only difference between fermions and bosons is the factor of $(-1)^{M_1-1}$ in the right-hand side in Eq.~\eqref{eq_SUNCS1}, hence is included into the quantum number in the charge sector $I_j^{(1)}$ as shown below. Then, by taking the logarithm, we obtain the transcendental ABA equations as
\begin{gather}
k_j^{(1)}L=2\pi I_j^{(1)}+\sum_m^{M_2}\theta(k_m^{(2)}-k_j^{(1)})+\tilde \lambda^\prime\sum_l^{M_1}\theta(k_j^{(1)}-k_l^{(1)}),\quad(1\le j\le M_1),\label{eq_SUNCS1log}\\
\sum_l^{M_\sigma}\theta(k_m^{(\sigma)}-k_l^{(\sigma)})+2\pi I_m^{(\sigma)}=\sum_j^{M_{\sigma-1}}\theta(k_m^{(\sigma)}-k_j^{(\sigma-1)})+\sum_q^{M_{\sigma+1}}\theta(k_m^{(\sigma)}-k_q^{(\sigma+1)}),\quad(1\le m\le M_\sigma,~2\le\sigma\le N-1),\label{eq_SUNCS2log}\\
\sum_l^{M_{N}}\theta(k_q^{(N)}-k_l^{(N)})+2\pi I_q^{(N)}=\sum_j^{M_{N-1}}\theta(k_q^{(N)}-k_j^{(N-1)}),\quad(1\le q\le M_{N}),\label{eq_SUNCS3log}
\end{gather}
\end{widetext}
where the quantum number $I_j^{(1)}$ that characterizes a charge excitation is given by
\begin{align}
I_j^{(1)}=\frac{M_2}{2},\quad(\mathrm{mod}~1),\label{eq_Ijfermion1}
\end{align}
for fermions, and 
\begin{align}
I_j^{(1)}=\frac{M_1+M_2+1}{2},\quad(\mathrm{mod}~1),
\label{eq_Ijboson1}
\end{align}
for bosons, respectively. On the other hand, $N-1$ kinds of spin excitations ($2\le\sigma \le N$) are described by the same quantum numbers in the case of fermions and bosons;
\begin{align}
I_j^{(\sigma)}=\frac{M_{\sigma-1}+M_\sigma+M_{\sigma+1}+1}{2},\quad(\mathrm{mod}~1,~2\le\sigma \le N),
\label{eq_Ij2}
\end{align}
where $M_{N+1}=0$. The energy of the system is given in terms of the quasimomentum $k_j^{(1)}$ in the charge sector as $E=(1/2)\sum_j[k_j^{(1)}]^2$. Importantly, dissipation parameter $\tilde \lambda^\prime$ only appears in the charge sector in Eq.~\eqref{eq_SUNCS1} and Eq.~\eqref{eq_SUNCS1log} in the nested ABA equations. This fact is nothing but the protection of spin sectors by SU($N$) internal symmetry in NH quantum systems, and can be seen as a NH generalization of the spin-charge separation in 1D quantum many-body systems \cite{Nakagawa20, Yamamoto22}.

\subsection{Velocity of excitations}
In this subsection, we obtain the velocity of excitations, for which we generalize the distribution function and the dressed energy to NH quantum systems with SU($N$) symmetry. We first consider the ground state for general cases, where the chemical potential imbalance between different spin-$\sigma$ particles is included. In such ground states, quantum numbers $\{I_j^{(\sigma)}\}$ are the successive integers distributed symmetrically around the origin. As the phase shift in the NH SU($N$) CS model is a step function with a complex coefficient, the quasimomentum is equally spaced in each region $C_\tau$ between $\pm Q_\tau$ and $\pm Q_{\tau+1}$, where $|Q_1|\ge |Q_2|\ge\cdots\ge |Q_N|$ and $Q_{N+1}=0$ (see Fig.~\ref{fig_phaseshiftSUN}). As we see from Fig.~\ref{fig_phaseshiftSUN}, the quasimomentum $k_j^{(\sigma)}$ is distributed in the region $\sum_{\tau=\sigma}^N C_\tau$, where $\pm Q_\sigma$ denote the endpoints of the quasimomentum distribution of $k_j^{(\sigma)}$. Accordingly, the distribution function in the thermodynamic limit $\rho_\sigma(k)$ introduced for the quasimomentum $k_j^{(\sigma)}$ takes a constant complex value in each region $C_\tau$, and is given by
\begin{align}
\rho_\sigma(k)=\frac{\tau-\sigma+1}{2\pi(1+\tau\tilde\lambda^\prime)},\quad (\tau\ge\sigma,~\sigma=1,2,\cdots N).
\label{eq_distributionSUN}
\end{align}
By using the distribution function \eqref{eq_distributionSUN}, the particle density with spin $\sigma$ is obtained as
\begin{align}
n_\sigma=\frac{N_\sigma}{L}=\int_{\sum_{\tau=\sigma}^N C_\tau}\rho_\sigma(k)dk-\int_{\sum_{\tau=\sigma+1}^N C_\tau}\rho_{\sigma+1}(k)dk,
\label{eq_particle_densitySUN}
\end{align}
where we have assumed $n_1\ge n_2\ge\cdots\ge n_N$, and the endpoint of each quasipaticle distribution $Q_\sigma$ is given by the following set of equations;
\begin{gather}
Q_N=\pi n_N(1+N\tilde\lambda^\prime),\\
Q_{N-1}-Q_N=\pi(n_{N-1}-n_N)(1+(N-1)\tilde\lambda^\prime),\\
~\vdots\notag\\
Q_1-Q_2=\pi(n_1-n_2)(1+\tilde\lambda^\prime),
\end{gather}
which become complex reflecting the quasimomentum distribution in the complex plane.

Now, we concentrate on the SU($N$) singlet ground state, where $Q_1=\cdots=Q_N=\pi n(1+N\tilde\lambda^\prime)/N$ with the total density $n=M/L$. In this case, only the region $C_N$ exists and the quasimomentum $k_j^{(\sigma)}$ is distributed in the region $C_N$ depicted by the straight line (see Fig.~\ref{fig_phaseshiftSUN}). By inserting a hole into the ground-state distribution of the quasimomentum, the dressed energy is calculated with the help of the nested ABA equations \eqref{eq_SUNCS1log}--\eqref{eq_SUNCS3log} as
\begin{widetext}
\begin{gather}
\epsilon_1(k)=\frac{1}{2}k^2-\tilde \mu +\int_{-Q_N}^{Q_N}\delta(k-k^\prime)\epsilon_2(k^\prime)dk^\prime-\tilde \lambda^\prime\int_{-Q_N}^{Q_N}\delta(k-k^\prime)\epsilon_1(k^\prime)dk^\prime,\\
\epsilon_\sigma(k)=\sum_{q=-1,0,1}(-1)^{q+1}\int_{-Q_N}^{Q_N}\delta(k-k^\prime)\epsilon_{\sigma+q}(k^\prime)dk^\prime,\quad(\sigma=2,3,\cdots,N-1),\\
\epsilon_N(k)=\sum_{q=-1,0}(-1)^{q+1}\int_{-Q_N}^{Q_N}\delta(k-k^\prime)\epsilon_{N+q}(k^\prime)dk^\prime,
\end{gather}
\end{widetext}
where the complex chemical potential $\tilde \mu$ is obtained by using $\epsilon_\sigma(\pm Q_N)=0$ as $\tilde \mu =\pi^2n^2(1+N\tilde \lambda^\prime)^2/2N^2$. Finally, we arrive at the velocity of excitations for the SU($N$) singlet ground state as
\begin{align}
\tilde v_\sigma=\frac{\epsilon_\sigma^\prime(Q_N)}{2\pi\rho_\sigma(Q_N)}=\frac{\pi n}{N}(1+N\tilde\lambda^\prime)\equiv \tilde v,
\end{align}
where $\epsilon_\sigma^\prime(k)$ denotes the derivative of $\epsilon_\sigma(k)$ with respect to $k$. Here, $\tilde v_1$ characterizes the charge velocity and $\tilde v_\sigma$ ($\sigma=2,3,\cdots,N$) describes the $N-1$ kinds of spin velocity. We note that the velocity for the charge excitation $\tilde v_1$ and that for the spin excitations $\tilde v_\sigma$ ($\sigma=2,3,\cdots,N$) are the same in long-range interacting systems \eqref{eq_HeffSUN}. It is worth noting that both the charge velocity and the spin velocity are affected by dissipation through $\tilde \lambda^\prime$ as a result of the nested equations of ABA.

\subsection{Finite-size spectrum}
In this subsection, we obtain the finite-size spectrum that leads to the universal scaling relations for dissipative TL liquids with SU($N$) spin symmetry in NH quantum critical systems. In Hermitian systems, SU($N$) CS model has been shown to describe the universal properties of TL liquids with SU($N$) spin symmetry as an ideal gas, e.g., SU(2) CS model successfully describes the SU(2) spin symmetric TL liquids, which characterize the critical properties of Hubbard models \cite{Kawakami92c, Kawakami93a, Kawakami90A, Schulz90, Frahm90}. The excitation energy in a finite system for the NH SU($N$) CS model can be obtained by using the nested ABA equations \eqref{eq_SUNCS1log}--\eqref{eq_SUNCS3log} \cite{Kawakami92b}. This method is useful to understand conformal properties stemming from U(1)$\times$SU($N$) symmetry underlying in the ABA equations as we see below. On the other hand, the finite-size spectrum for the Hermitian SU($N$) CS model was originally obtained by Ha and Haldane \cite{Ha92} with the use of the bosonization method, which also embodies an ideal-gas description. This method is beneficial to interpret the particle picture and the origin of associated selection rules for Fermi and Bose statistics. To understand the effect of dissipation on each particle more clearly, we start from the particle picture introduced in Ref.~\cite{Ha92} and then transform it to the ABA picture underlying in Eqs.~\eqref{eq_SUNCS1log}--\eqref{eq_SUNCS3log}.

By constructing excited states over the SU($N$) singlet ground state in Eq.~\eqref{eq_JastrowSUN} systematically, we end up with the finite-size spectrum, which is given by analytically continuing the one obtained in Ref.~\cite{Ha92} to the case of the NH SU($N$) CS model. The result is written down as
\begin{gather}
\Delta E = \frac{\pi \tilde v}{L}\sum_{\sigma \sigma^\prime}\left[\frac{1}{2}A_{\sigma\sigma^\prime}\Delta M_\sigma \Delta M_{\sigma^\prime}+2A^{-1}_{\sigma\sigma^\prime}\Delta J_\sigma \Delta J_{\sigma^\prime}\right],\label{eq_DeltaESUN}
\end{gather}
where we have to pay careful attention to the fact that not only the interaction parameter $\tilde \lambda^\prime$ but also the velocity of excitations $\tilde v$ becomes complex-valued as a result of dissipation \cite{Yamamoto22}. We also note that particle-hole excitation terms that form conformal towers inherent in the underlying Virasoro algebra are omitted in Eq.~\eqref{eq_DeltaESUN} because they do not affect the critical exponents of leading order that we consider in the next subsection. In Eq.~\eqref{eq_DeltaESUN}, the coefficient $A_{\sigma\sigma^\prime}=\tilde\lambda^\prime+\delta_{\sigma\sigma^\prime}$ is the matrix element of the $N\times N$ matrix $\bm A$ given by
\begin{gather}
\bm A=\left(\begin{matrix}
\tilde\lambda^\prime+1&\tilde\lambda^\prime&\tilde\lambda^\prime&\cdots\\
\tilde\lambda^\prime&\tilde\lambda^\prime+1&\tilde\lambda^\prime&\\
\tilde\lambda^\prime&\tilde\lambda^\prime&\ddots&\\
\vdots&&&\\
\end{matrix}\right),\label{eq_matrixA}
\end{gather}
and $A^{-1}_{\sigma\sigma^\prime}=-{\tilde \lambda^\prime}/(1+N\tilde\lambda^\prime)+\delta_{\sigma\sigma^\prime}$ is that of the inverse matrix $\bm A^{-1}$, which is given by
\begin{gather}
\bm A^{-1}=\left(\begin{matrix}
\chi+1&\chi&\chi&\cdots\\
\chi&\chi+1&\chi&\\
\chi&\chi&\ddots&\\
\vdots&&&\\
\end{matrix}\right),\label{eq_matrixAinverse}
\end{gather}
where we have introduced $\chi=-\tilde \lambda^\prime/(1+N\tilde\lambda^\prime)$. We note that $\Delta M_\sigma$ and $\Delta J_\sigma$ are the vector elements of the $N\times1$ column vectors that characterize two types of excitations given by $\Delta \bm M$ and $\Delta \bm J$, respectively. Here, $\Delta M_\sigma$ denotes the particle-number excitations, and $\Delta J_\sigma$ describes the excitations accompanied by the large momentum transfer $P_\sigma$ for a spin-$\sigma$ particle, which is given by
\begin{align}
P_\sigma=2k_F\Delta J_\sigma,
\label{eq_Pelectron}
\end{align}
where the Fermi momentum $k_F=\pi n / N$ is generalized to the SU($N$) singlet ground state.

As we see in Eq.~\eqref{eq_matrixA}, dissipation affects all the elements of the matrix $\bm A$ equally via the complex-valued interaction parameter $\tilde \lambda^\prime$. We emphasize that the difference of statistics between fermions and bosons are incorporated into the selection rules of quantum numbers, which are given by
\begin{gather}
\Delta M_\sigma:~\mathrm{integer},\label{eq_deltaMsigma}\\
\Delta J_\sigma=\frac{\Delta M_\sigma}{2}~(\mathrm{mod}~1),
\end{gather}
for fermions, and
\begin{gather}
\Delta M_\sigma:~\mathrm{integer},\\
\Delta J_\sigma:~\mathrm{integer},\label{eq_deltaJsigma}
\end{gather}
for bosons, respectively. We see that the selection rules for each spin-$\sigma$ particle \eqref{eq_deltaMsigma}--\eqref{eq_deltaJsigma} are the same as those in the single component cases \eqref{eq_deltaM}--\eqref{eq_deltaD}.

The basis of the matrices used in Eqs.~\eqref{eq_matrixA} and \eqref{eq_matrixAinverse} are the one which was introduced by Ha and Haldane \cite{Ha92}, and is different from the one used in the ABA equations \eqref{eq_SUNCS1log}--\eqref{eq_SUNCS3log}. The basis transformation between them is conducted by using the following $N\times N$ matrix $\bm \alpha$ \cite{Kawakami93b}
\begin{gather}
\bm\alpha=\left(\begin{matrix}
1&-1&&&\\
&1&-1&\mbox{\Large${\mathbf{0}}$}&\\
&&\ddots&\ddots&\\
&\mbox{\Large${\mathbf{0}}$}&&\ddots&-1\\
&&&&1\\
\end{matrix}\right),\label{eq_alpha}
\end{gather}
which gives the inverse matrix as
\begin{gather}
\bm\alpha^{-1}=\left(\begin{matrix}
1&1&1&\cdots&1\\
&1&1&\cdots&1\\
&&1&\cdots&1\\
&\mbox{\Large$\mathbf 0$}&&\ddots&1\\
&&&&1\\
\end{matrix}\right).\label{eq_alphainverse}
\end{gather}
Then, with the help of Eq.~\eqref{eq_alpha}, the excitation energy \eqref{eq_DeltaESUN} is rewritten as $\Delta E = (\pi \tilde v / {L})(\bm m^t\bm D\bm m/2+2\bm d^t \bm D^{-1}\bm d)$, where the matrix $\bm A$ is transformed into the matrix $\bm D$, which is calculated as
\begin{gather}
\bm D\equiv\bm\alpha^t\bm A \bm\alpha
=\left(\begin{matrix}
\tilde\lambda^\prime+1&-1&&&\\
-1&2&-1&\mbox{\Large${\mathbf{0}}$}&\\
&-1&2&\ddots&\\
&\mbox{\Large${\mathbf{0}}$}&\ddots&\ddots&-1\\
&&&-1&2\\
\end{matrix}\right).\label{eq_cartan}
\end{gather}
Similarly, with the use of Eq.~\eqref{eq_alphainverse}, the inverse matrix $\bm D^{-1}$ is written as follows;
\begin{widetext}
\begin{gather}
\bm D^{-1}\equiv\bm \alpha^{-1}\bm A^{-1} (\bm \alpha^t)^{-1}
=\frac{1}{1+N\tilde\lambda^\prime}\left(\begin{matrix}
N&N-1&N-2&&\cdots&1\\
N-1&(N-1)(\tilde\lambda^\prime+1)&(N-2)(\tilde\lambda^\prime+1)&&\cdots&\tilde\lambda^\prime+1\\
N-2&(N-2)(\tilde\lambda^\prime+1)&(N-2)(2\tilde\lambda^\prime+1)&&&\vdots\\
\vdots&\vdots&\vdots&&&(N-2)\tilde\lambda^\prime+1\\
1&\tilde\lambda^\prime+1&2\tilde\lambda^\prime+1&&\cdots&(N-1)\tilde\lambda^\prime+1\\
\end{matrix}\right).
\end{gather}
\end{widetext}
Accordingly, the basis transformations for the quantum number vectors $\Delta \bm M$ and $\Delta\bm J$ are given by
\begin{align}
\bm m &=\bm \alpha^{-1}\Delta \bm M,\\
\bm d &=\bm \alpha^t\Delta \bm J,
\end{align}
where the quantum numbers $m_1$ and $d_1$ characterize excitations in the charge sector with the Fermi point $Nk_F$, and $m_\sigma$ and $d_\sigma$ ($2\le\sigma\le N$) describe excitations in the spin sectors with the Fermi point $(N-\sigma+1)k_F$. In detail, $m_\sigma$ denotes the change of the number of charge particles for $\sigma=1$, and that of spinons for $2\le\sigma\le N$. Accordingly, $d_\sigma$ carries the large momentum $P_\sigma^\prime$ associated with charge ($\sigma=1$) or spin ($2\le\sigma\le N$) excitations as follows;
\begin{align}
P_\sigma^\prime=2(N-\sigma+1)k_Fd_\sigma.
\label{eq_Pspinon}
\end{align}
By using the quantum numbers $m_\sigma$ and $d_\sigma$, the selection rules are rewritten as 
\begin{gather}
m_\sigma:~\mathrm{integer},\label{eq_msigma}\\
d_\sigma=\frac{m_{\sigma-1}+m_{\sigma+1}}{2}~(\mathrm{mod}~1),
\end{gather}
with $m_0=m_1$ and $m_{N+1}=0$ for fermions, and
\begin{gather}
m_\sigma:~\mathrm{integer},\\
d_\sigma:~\mathrm{integer},\label{eq_dsigma}
\end{gather}
for bosons, respectively. The selection rules for $m_\sigma$ and $d_\sigma$ given in Eqs.~\eqref{eq_msigma}--\eqref{eq_dsigma} are rather complicated than those for $\Delta M_\sigma$ and $\Delta J_\sigma$ given in Eqs.~\eqref{eq_deltaMsigma}--\eqref{eq_deltaJsigma}, but are easily obtained by using the basis transformation given by the matrix $\bm \alpha$ in Eq.~\eqref{eq_alpha}.

Notably, the matrix $\bm D$ in Eq.~\eqref{eq_cartan} reflects the SU($N$) internal spin symmetry underlying in the present dissipative TL liquids. $(N-1)\times(N-1)$ matrix obtained by deleting the first column and the first row in $\bm D$ is called the SU($N$) Cartan matrix, which characterizes $N-1$ kinds of spin excitations with $c=N-1$ level-1 SU($N$) Kac-Moody algebra \cite{Kac84, Knizhnik84, Witten84, Goddard86, Polyakov83}. On the other hand, we see in Eq.~\eqref{eq_cartan} that the effect of dissipation through the complex parameter $\tilde \lambda^\prime$ only appears in the charge sector, which is characterized by a complex generalization of $c=1$ U(1) Gaussian CFT  \cite{Yamamoto22} as in the case without internal spin symmetry discussed in Sec.~\ref{sec_dissipativeTL}.

\subsection{Universal scaling relations}
Finally, we obtain the critical exponents which provide universal properties of dissipative TL liquids with SU($N$) spin symmetry. From the excitation energy spectrum \eqref{eq_DeltaESUN} and the basis transformation given in Eq.~\eqref{eq_cartan}, the general formula of the (complex) critical exponents that express dissipative TL liquids with SU($N$) spin symmetry are read off as \cite{Izergin89}
\begin{align}
\tilde\eta&=\frac{1}{2}\Delta\bm M^t\bm A\Delta \bm M+2\Delta \bm J^t \bm A^{-1}\Delta \bm J\label{eq_electron_basis}\\
&=\frac{1}{2}\bm m^t\bm D\bm m+2\bm d^t \bm D^{-1}\bm d,\label{eq_spinon_holon_basis}
\end{align}
where the first line \eqref{eq_electron_basis} corresponds to the basis introduced by Ha and Haldane \cite{Ha92}, and the second line \eqref{eq_spinon_holon_basis} corresponds to the one used in the ABA equations \eqref{eq_SUNCS1log}--\eqref{eq_SUNCS3log}. It is noted that $\tilde \eta$ is the critical exponent of biorthogonal correlation functions in NH systems \cite{Yamamoto22}, hence is a complex value. In the following, we obtain the scaling formula for both biorthogonal correlation functions and right-state correlation functions in NH systems based on the formula of $\tilde \eta$ given in Eqs.~\eqref{eq_electron_basis} and \eqref{eq_spinon_holon_basis}.

\subsubsection{Critical exponents for biorthogonal correlation functions}
We first calculate the critical exponents for biorthogonal correlation functions. The exponents of biorthogonal correlators become complex, and are directly given by the formula $\tilde \eta$ given in Eqs.~\eqref{eq_electron_basis} and \eqref{eq_spinon_holon_basis}. First, the long-distance behavior of the biorthogonal charge-density correlator is given by
\begin{align}
{}_L\langle\rho(x)\rho(0)\rangle_R\simeq\sum_{j=1}^N\tilde A_j \cos[2(N-j+1)k_Fx]x^{-\tilde \beta_j}+\frac{\tilde A_0}{x^2},\label{eq_biorthogonaldensity}
\end{align}
which has the same form both in the case of fermions and bosons. Here, the exponent $\tilde \beta_j$ denotes the $2(N-j+1)k_F$ oscillating part $(1\le j \le N)$ of the biorthogonal charge-density correlator \eqref{eq_biorthogonaldensity}, and $\tilde A_j$ is a complex-valued correlation amplitude. As Eq.~\eqref{eq_biorthogonaldensity} conserves the particle number, the selection rules for $\Delta \bm M$ and $\bm m$ for this type of excitations reduce to
\begin{align}
\Delta \bm M =\bm m =\bm 0,
\end{align}
which brings about the same form of the charge density correlators for both fermions and bosons. For the excitations that convey the large momentum transfer given in Eqs.~\eqref{eq_Pelectron} and \eqref{eq_Pspinon}, the selection rules corresponding to the $2(N-j+1)k_F$ oscillations $(1\le j \le N)$ associated with the momentum transfer $2Nk_F,~2(N-1)k_F,\cdots,2k_F$ are given by
\begin{align}
&2Nk_F~\mathrm{oscillation},\notag\\
&\quad\Delta \bm J =(1,1,\cdots,1)^t, \quad\bm d =(1,0,0,\cdots,0)^t,\label{eq_2Nosc}\\
\notag\\
&2(N-1)k_F~\mathrm{oscillation},\notag\\
&\quad\Delta \bm J =(0,1,\cdots,1)^t, \quad\bm d =(0,1,0,\cdots,0)^t,\\
&\quad\vdots\notag\\
&2k_F~\mathrm{oscillation},\notag\\
&\quad\Delta \bm J =(0,\cdots,0,1)^t,\quad\bm d =(0,0,\cdots,0,1)^t,\label{eq_2osc}
\end{align}
respectively. Equations \eqref{eq_2Nosc}--\eqref{eq_2osc} provide the critical exponent $\tilde \beta_j$ corresponding to the $2(N-j+1)k_F$ oscillating piece $(1\le j \le N)$ as
\begin{align}
\tilde \beta_j=\frac{2(N-j+1)(j-1)}{N}+\frac{2(N-j+1)^2}{N}\tilde K_\rho,
\label{eq_beta_tilde}
\end{align}
where the complex-valued TL parameter $\tilde K_\rho$ is given by $\tilde K_\rho=1/(1+N\tilde \lambda^\prime)$ with the use of Eq.~\eqref{eq_complexTLparameterSUN}. Notably, from Eq.~\eqref{eq_beta_tilde}, we see that the effect of dissipation is included only in the charge excitation described by $\tilde K_\rho$.

On the other hand, fermion correlators and boson correlators are different from each other. The long-distance behavior of the biorthogonal fermion correlator, which accompanies the $k_F$ oscillation, is written as
\begin{align}
{}_L\langle c_\sigma^\dag(x)c_\sigma(0)\rangle_R\simeq \tilde C_1 \cos(k_Fx)x^{-\tilde \eta_F},
\label{eq_biorthogonalfermion}
\end{align}
where $\tilde C_1$ is a complex-valued correlation amplitude. It should be noted that the biorthogonal fermion correlator \eqref{eq_biorthogonalfermion} is independent of spin indices for the SU($N$) singlet ground state. As the correlation function \eqref{eq_biorthogonalfermion} changes the particle number by one, the selection rules for the biorthogonal fermion correlator \eqref{eq_biorthogonalfermion} are given by
\begin{align}
&\sigma=1\notag\\
&\quad\Delta \bm M=(1,0,\cdots,0)^t,\quad\bm m=(1,0,\cdots,0)^t,\label{eq_rulefermion1}\\
&\quad\Delta \bm J =(\frac{1}{2},0,\cdots,0)^t,\quad \bm d =(\frac{1}{2},-\frac{1}{2},0,\cdots,0)^t,\\
\notag\\
&\sigma=2\notag\\
&\quad\Delta \bm M=(0,1,0,\cdots,0)^t,\quad\bm m=(1,1,0,\cdots,0)^t,\\
&\quad\Delta \bm J =(0,\frac{1}{2},0,\cdots,0)^t,\quad \bm d =(0,\frac{1}{2},-\frac{1}{2},0,\cdots,0)^t,\\
&\quad\vdots\notag\\
&\sigma=N\notag\\
&\quad\Delta \bm M=(0,\cdots,0,1)^t,\quad\bm m=(1,1,\cdots,1)^t,\\
&\quad\Delta \bm J =(0,\cdots,0,\frac{1}{2})^t,\quad \bm d =(0,\cdots,0,\frac{1}{2})^t,\label{eq_rulefermionN}
\end{align}
all of which lead to the same critical exponent $\tilde \eta_F$ as
\begin{align}
\tilde \eta_F=\frac{N-1}{N}+\frac{1}{2N\tilde K_\rho}+\frac{\tilde K_\rho}{2N}.
\label{eq_etaF_tilde}
\end{align}
Again, we see from Eq.~\eqref{eq_etaF_tilde} that dissipation only affects the exponent of the charge excitation characterized by $\tilde K_\rho$, and the exponent stemming from spin excitations, $(N-1)/N$, is fixed to be real as a result of the underlying SU($N$) symmetry. 

The long-distance behavior of the biorthogonal boson correlator, which does not accompany any oscillations, are given by
\begin{align}
{}_L\langle b_\sigma^\dag(x)b_\sigma(0)\rangle_R\simeq \tilde B_1x^{-\tilde \eta_B},
\label{eq_biorthogonalboson}
\end{align}
where $\tilde B_1$ is a complex-valued correlation amplitude. We note that the boson correlator in Eq.~\eqref{eq_biorthogonalboson} is independent of spin indices for the SU($N$) singlet ground state as in the fermion case \eqref{eq_biorthogonalfermion}. As the leading part of the biorthogonal boson correlator \eqref{eq_biorthogonalboson} does not carry the large momentum from the left Fermi point to the right one, the selection rules for $\Delta\bm J$ and $\bm d$ read
\begin{align}
\Delta \bm J=\bm d=\bm 0.
\end{align}
The boson correlator \eqref{eq_biorthogonalboson} changes the number of particles by one as in the fermion case \eqref{eq_biorthogonalfermion}, hence the selection rules for $\Delta \bm M$ and $\bm m$ are the same as Eqs.~\eqref{eq_rulefermion1}--\eqref{eq_rulefermionN}. Then, we obtain the critical exponent $\tilde \eta_B$ as 
\begin{align}
\tilde \eta_B=\frac{N-1}{2N}+\frac{1}{2N\tilde K_\rho},
\label{eq_etaB_tilde}
\end{align}
where the piece proportional to the TL parameter does not appear in contrast to the one in the fermion case \eqref{eq_etaF_tilde}. Importantly, the critical exponents of the biorthogonal correlation functions \eqref{eq_beta_tilde}, \eqref{eq_etaF_tilde}, and \eqref{eq_etaB_tilde} provide the universal scaling relations for dissipative TL liquids with SU($N$) spin symmetry.

\subsubsection{Critical exponents for right-state correlation functions}
Next, we analyze the right-state correlation functions, which are experimentally relevant in ultracold atoms. We emphasize that the exponents of right-state correlators are real because right-state correlation functions are obtained as a standard quantum-mechanical expectation value for the ground state \cite{Ashida16, Yamamoto22}.

The critical exponents of the right-state correlation functions are obtained from Eqs.~\eqref{eq_beta_tilde}, \eqref{eq_etaF_tilde}, and \eqref{eq_etaB_tilde} by replacing $\tilde K_\rho$ with $K_\rho^\phi$, and $1/\tilde K_\rho$ with $1/K_\rho^\theta$, respectively. This simple transformation comes from the fact that dissipation only affects the charge excitation characterized by the complex-valued TL parameter $\tilde K_\rho$. As $\tilde K_\rho$ describes the charge degrees of freedom, the relations among the TL parameters $\tilde K_\rho$, $K_\rho^\phi$, and $K_\rho^\theta$ given in Eqs.~\eqref{eq_Kphi} and \eqref{eq_Ktheta} hold for dissipative TL liquids both with and without internal symmetry \cite{Yamamoto22}. As a result, two types of real TL parameters $K_\rho^\phi$ and $K_\rho^\theta$ describe the universal properties of right-state correlation functions through the general formula \eqref{eq_Kphi} and \eqref{eq_Ktheta} in dissipative TL liquids with SU($N$) spin symmetry. Therefore, in terms of $K_\rho^\phi$ and $K_\rho^\theta$, we obtain the universal scaling relations for $\beta_j$, $\eta_F$, and $\eta_B$ in the right-state correlation functions as Eqs.~\eqref{eq_beta_SUN}, \eqref{eq_etaFSUN}, and \eqref{eq_etaBSUN} as summarized in the main results in Sec.~\ref{sec_main_results}. These exponents characterize the universal behavior of dissipative TL liquids with SU($N$) spin symmetry, and are relevant to experiments.

\begin{figure}[t]
\centering
\includegraphics[width=8.5cm]{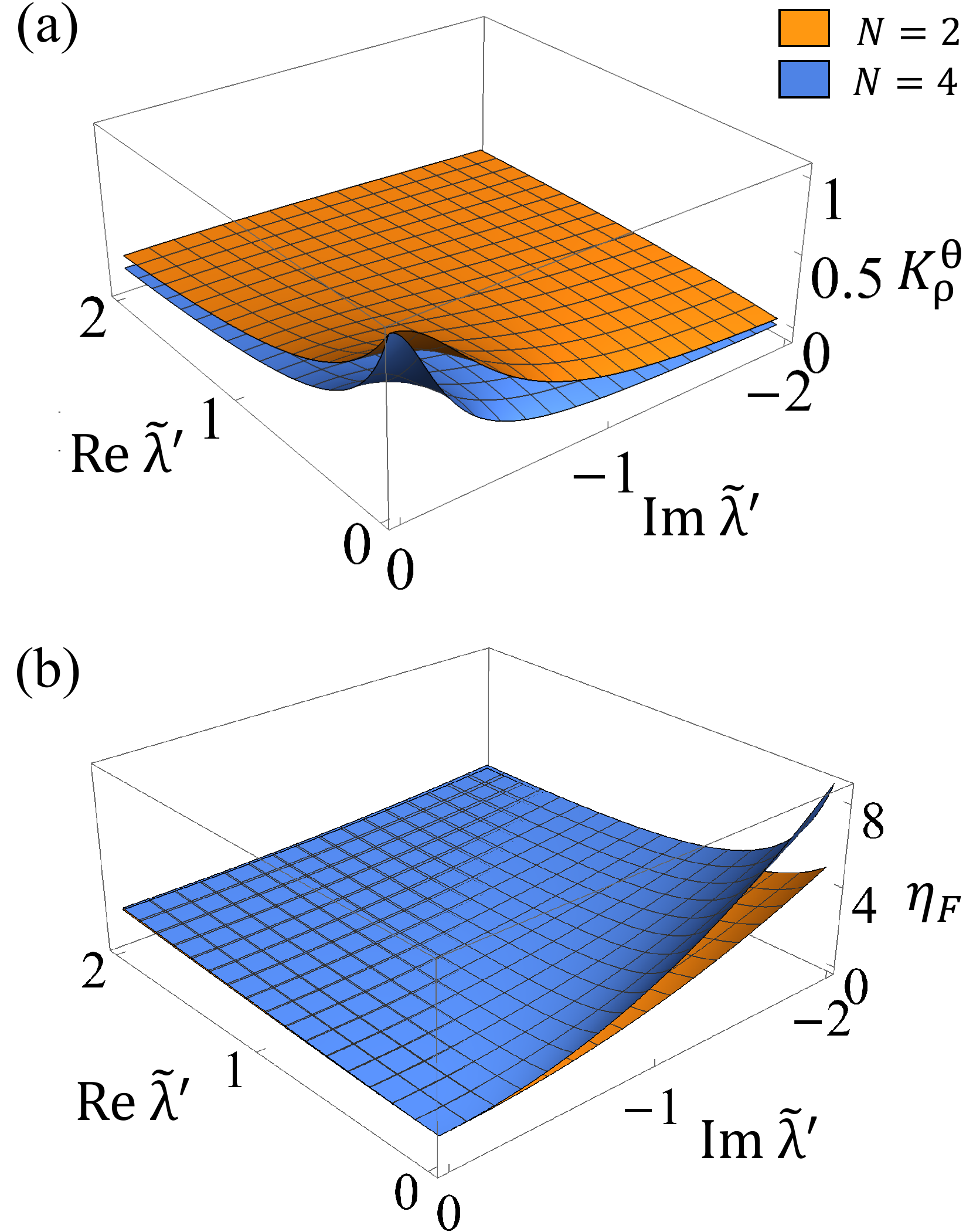}
\caption{Critical exponents (a) $K_\rho^\theta$ and (b) $\eta_F$ as a function of the complex-valued interaction parameter $\tilde \lambda^\prime$ for $N=2$ (orange) and $N=4$ (blue) in the NH SU($N$) CS model \eqref{eq_HeffSUN}. $\mathrm{Re}\tilde \lambda^\prime$ and $\mathrm{Im}\tilde \lambda^\prime$ denote a (real-valued) interaction, and dissipation, respectively. $K_\rho^\theta$ and $\eta_F$ do not depend on the sign of $\mathrm{Im}\tilde \lambda^\prime$. }
\label{fig_CriticalExponents}
\end{figure}

In Fig.~\ref{fig_CriticalExponents}, we plot the critical exponents $K_\rho^\theta$ given in Eq.~\eqref{eq_Ktheta} and $\eta_F$ given in Eq.~\eqref{eq_etaFSUN} in long-range interacting SU($N$) symmetric systems \eqref{eq_HeffSUN} for the case of $N=2$ and $N=4$. We note that $K_\rho^\phi$ is not affected by dissipation in the present case due to an artifact of long-range interactions. From Fig.~\ref{fig_CriticalExponents}(a), we see that the TL parameter $K_\rho^\theta$ is suppressed as dissipation $\mathrm{Im}\tilde \lambda^\prime$ is increased. This is a similar behavior obtained in the NH Lieb-Liniger model \cite{Ashida16}, where the TL parameter is suppressed due to dissipation as a result of the continuous quantum Zeno effect \cite{Syassen08, Garcia09, Rey14, Yan13, Yamamoto19, Yamamoto21}. As we deal with the continuum model, the increase in TL parameters as a result of the Umklapp scattering reported in lattice models is not seen \cite{Yamamoto22}. Accordingly, the behavior of $\eta_F$ shown in Fig.~\ref{fig_CriticalExponents}(b) reflects the suppression of $K_\rho^\theta$ via dissipation $\mathrm{Im}\tilde \lambda^\prime$, and shows the enhancement due to dissipation. The enhancement in $\eta_F$ gives the suppression of the fermion correlator \eqref{eq_correlator_fermion}, the behavior of which is consistent with the one obtained in the NH Lieb-Liniger model \cite{Ashida16}. In Fig.~\ref{fig_CriticalExponents}(b), we see that the enhancement of $\eta_F$ due to dissipation $\mathrm{Im}\tilde \lambda^\prime$ becomes significant in the limit of free fermions, and the difference of $\eta_F$ induced by dissipation between $N=2$ and $N=4$ also becomes significant in this limit. We note that almost the same behavior as that of $\eta_F$ is obtained for $\eta_B$ given in Eq.~\eqref{eq_etaBSUN}.


\section{Conclusions}
\label{sec_discussion}
We have explored the universal properties of dissipative TL liquids with SU($N$) spin symmetry in one dimension based on a complex generalization of Haldane's ideal-gas description, which is realized by the SU($N$) Calogero-Sutherland model.  As the main results, we have obtained the universal scaling relations for dissipative TL liquids with SU($N$) spin symmetry for both fermions and bosons. We have demonstrated that the spectrum of dissipative TL liquids with SU($N$) spin symmetry is described by the sum of one charge mode characterized by a complex generalization of $c=1$ U(1) Gaussian CFT, and $N-1$ spin modes characterized by level-$1$ SU($N$) Kac-Moody algebra with the conformal anomaly $c=N-1$. We have elucidated that dissipation only affects the charge sector in the correlation exponents by deriving ABA solutions and using CFT in NH quantum critical many-body systems. The scaling relations obtained in the present work are relevant to a wide variety of cold-atom experiments in NH quantum critical systems with SU($N$) symmetry, e.g., ultracold alkaline-earth-like atom ${}^{173}$Yb is a promising candidate, where the Fermi-Hubbard model is successfully loaded into an optical lattice, and dissipation is introduced by using photoassociation techniques \cite{Honda22}. As a useful tool to postselect special measurement outcomes that realize NH quantum many-body systems, quantum-gas microscopy can be utilized \cite{Ashida16, Ashida17, Nakagawa20, Yamamoto22} to observe dissipative TL liquids with SU($N$) spin symmetry.

There remains an interesting question how the present results for dissipative TL liquids are related to the critical behavior studied in open quantum systems \cite{Honing12, Bacsi20}. Critical exponents of correlation functions and correlation length are essential to investigate the microscopic origin of quantum phase transitions. As critical exponents in open quantum systems have been investigated by using the framework of Lindblad master equations \cite{Honing12, Bacsi20}, it is important to clarify the universality class which lies in dissipative quantum critical systems from the perspective of CFT. As for the relations between NH quantum systems and Markovian open quantum systems, it is known that the spectrum of the Liouvillian is obtained from that of the NH Hamiltonian when the system follows loss-only dynamics \cite{Torres14, Buca20, Nakagawa21}. This may benefit us for finding out universal relations of critical exponents between NH quantum systems and Markovian open quantum systems. Moreover, in view of recent advancement of CFT description in NH quantum systems \cite{Couvreur17, Ryu20, Kawabata22} and measurement-induced entanglement dynamics \cite{Li19, Ludwig20, Chen20, Ludwig21}, it is interesting to investigate how CFT in NH quantum many-body systems is related to the one in measurement-induced dynamics.

Dissipation makes the theory nonunitary and the entire spectrum of the NH quantum many-body systems can be complicated. However, low-energy physics shows universal properties in NH quantum systems with internal degrees of freedom, which leads to unconventional quantum critical phenomena characterized by a complex extension of CFT. Since systems with inverse-square long-range interactions are known to be closely related to the fractional quantum Hall effect \cite{Haldane91fra, Kawakami93}, characterization of NH fractional quantum Hall effects \cite{Yoshida19, Yoshida20} by using the present framework remains a future research subject. As ultracold mixtures with SU($N$)$\times$SU($N^\prime$) symmetry have been realized \cite{Taie10}, it is of interest to extend our theory to more general multicomponent systems beyond SU($N$) symmetry. Our results in this paper will certainly stimulate further study on multicomponent extension of dissipative TL liquids in open quantum systems \cite{Buchhold15, Buchhold21, Ashida16, Ashida17, Bernier20, Dora20, Bacsi20, Bacsi22, Bacsi20L, Moca21, Dora21, Dora22}.

\begin{acknowledgments}
This work was supported by KAKENHI (Grants No.\ JP18H01140 and No.\ JP19H01838). K.Y. was supported by WISE Program, MEXT and JSPS KAKENHI Grant-in-Aid for JSPS fellows Grant No.\ JP20J21318.
\end{acknowledgments}

\appendix

\section{Ground-state wave function}
\label{sec_Jastrow}
In this appendix, we give a proof that a complex generalization of the wave function of the Jastrow form \eqref{eq_Jastrow} provides the eigenstate of the effective Hamiltonian \eqref{eq_Heff} with the complex-valued ground-state energy $E_0$ given in Eq.~\eqref{eq_groundenergy}. Here, we consider the case of bosons for simplicity, and that of fermions can be proved similarly by taking the anticommutation relations of fermions into account. We calculate the term $-\sum_j \partial^2\Psi_g/{\partial x_j^2}$ by using the identity
\begin{align}
\frac{\partial \log \Psi_g}{\partial x_j}=\frac{1}{\Psi_g}\frac{\partial\Psi_g}{\partial x_j}.
\end{align}
Since ${\partial \log \Psi_g}/{\partial x_j}$ is calculated as
\begin{align}
\frac{\partial \log \Psi_g}{\partial x_j}=\frac{\pi\tilde \lambda}{L}\sum_{i(\neq j)}\cot\frac{\pi(x_j-x_i)}{L},
\end{align}
\begin{widetext}
we obtain
\begin{align}
-\frac{\partial^2}{\partial x_j^2}\Psi_g
&=-\frac{\pi^2\tilde \lambda}{L^2}\left[\tilde \lambda \left(\sum_{i(\neq j)}\cot\frac{\pi(x_j-x_i)}{L}\right)^2-\sum_{i(\neq j)}\frac{1}{\sin^2(\pi(x_j-x_i)/L)}\right]\Psi_g\notag\\
&=-\frac{\pi^2\tilde \lambda}{L^2}\Bigg(\sum_{i(\neq j)}\frac{\tilde \lambda-1}{\sin^2(\pi(x_j-x_i)/L)}- \tilde \lambda (N-1)+ 2\tilde \lambda\sum_{\langle i, k\rangle(\neq j)}\cot\frac{\pi(x_j-x_i)}{L}\cot\frac{\pi(x_j-x_k)}{L}\Bigg)\Psi_g,
\label{eq_psig_i}
\end{align}
where we have used the identity $\cot^2x={1}/{\sin^2x}-1$ in the second equality, $\langle i, k \rangle$ denotes the combination over the pair of $i$ and $k$ with $i\neq k$, and $\langle i, k \rangle(\neq j)$ means that both $i$ and $k$ are different from $j$. By taking the sum of Eq.~\eqref{eq_psig_i}, we get
\begin{align}
&-\sum_j\frac{\partial^2}{\partial x_j^2}\Psi_g\notag\\
&=\frac{\pi^2\tilde\lambda}{L^2}\Bigg[-2\sum_{i > j}\frac{\tilde\lambda-1}{\sin^2(\pi(x_i-x_j)/L)}+ \tilde \lambda N(N-1) -2\tilde \lambda\sum_{\langle i, j, k\rangle}\left(\cot\frac{\pi(x_j-x_i)}{L}\cot\frac{\pi(x_j-x_k)}{L}+(i\to j\to k)\right)\Bigg]\Psi_g,
\end{align}
where $\langle i, j, k \rangle$ denotes the combination over the pair of $\{i,j,k\}$ which are different from each other, and $(i\to j \to k)$ stands for two terms of cyclic permutations of the first term in the same parentheses. Finally, by using the identity
\begin{align}
\cot x \cot y + \cot y \cot z + \cot z \cot x =1,\quad(x+y+z=0),
\end{align}
we arrive at
\begin{align}
-\sum_j\frac{\partial^2}{\partial x_j^2}\Psi_g=\left(\frac{\pi^2N(N^2-1)\tilde \lambda^2}{3L^2}-\frac{2\pi^2\tilde \lambda (\tilde \lambda-1)}{L^2}\sum_{i>j}\frac{1}{\sin^2(\pi(x_i-x_j)/L)}\right)\Psi_g.
\label{eq_schrodinger}
\end{align}
One notices that the second term in the right-hand side of Eq.~\eqref{eq_schrodinger} has the same form as the interaction term \eqref{eq_interaction}. Therefore, by taking $2\tilde \lambda (\tilde \lambda -1) = \tilde g$, the wave function \eqref{eq_Jastrow} becomes the eigenstate with the ground state energy $E_0$ given in Eq.~\eqref{eq_groundenergy}.

\end{widetext}
\nocite{apsrev41Control}
\bibliographystyle{apsrev4-1}
\bibliography{NH_SU_N.bib}

\begin{thebibliography}{146}
\bibitem [{Foo()}]{Footnote}%
  \BibitemOpen
  \href@noop {} {}\bibinfo {note} {In the research field of high-energy
  physics, NH Calogero-Sutherland model with PT symmetry has been actively
  investigated (See reviews \cite{Ghosh12, Fring13} and references
  therein).}\BibitemShut {Stop}%
\end{thebibliography}

\end{document}